\documentclass[aps,twocolumn,
superscriptaddress,
footinbib,
prb]{revtex4-2}

\usepackage{amsmath,amssymb,bm}
\usepackage{mathrsfs}
\usepackage{MnSymbol}

\usepackage{graphicx}
\usepackage{epstopdf}
\usepackage{latexsym}
\usepackage[normalem]{ulem} 
\usepackage[caption=false]{subfig}
\usepackage{subfig}
\usepackage[usenames,dvipsnames]{color}
\usepackage{hyperref}
\usepackage{natbib}
\usepackage{xcolor}
\usepackage{physics}
\usepackage{multirow}
\usepackage{stackengine}
\usepackage{dsfont}
\usepackage{relsize}
\usepackage{float}

\usepackage{framed}

\usepackage[utf8]{inputenc} 

\renewcommand{\vec}{\boldsymbol}

\begin{document}

\title{Optimizing Design Choices for Neural Quantum States}

\date{\today}

\author{Moritz Reh}
\email{moritz.reh@kip.uni-heidelberg.de}
\affiliation{Kirchhoff-Institut f\"{u}r Physik, Universit\"{a}t Heidelberg, Im Neuenheimer Feld 227, 69120 Heidelberg, Germany}
\author{Markus Schmitt}
\affiliation{Forschungszentrum J\"ulich GmbH, Peter Gr\"unberg Institute, Quantum Control (PGI-8), 52425 J\"ulich, Germany}
\author{Martin G\"{a}rttner}
\affiliation{Kirchhoff-Institut f\"{u}r Physik, Universit\"{a}t Heidelberg, Im Neuenheimer Feld 227, 69120 Heidelberg, Germany}
\affiliation{Physikalisches Institut, Universit\"at Heidelberg, Im Neuenheimer Feld 226, 69120 Heidelberg, Germany}
\affiliation{Institut f\"ur Theoretische Physik, Ruprecht-Karls-Universit\"at Heidelberg, Philosophenweg 16, 69120 Heidelberg, Germany}

\begin{abstract}
Neural quantum states are a new family of variational ansätze for quantum-many body wave functions with advantageous properties in the notoriously challenging case of two spatial dimensions.
Since their introduction a wide variety of different network architectures has been employed to study paradigmatic models in quantum many-body physics with a particular focus on quantum spin models.
Nonetheless, many questions remain about the effect that the choice of architecture has on the performance on a given task. 
In this work, we present a unified comparison of a selection of popular network architectures and symmetrization schemes employed for ground state searches of prototypical spin Hamiltonians, namely the two-dimensional transverse-field Ising model and the $J_1$-$J_2$ model. In the presence of a non-trivial sign structure of the ground states, we find that the details of symmetrization crucially influence the performance. We describe this effect in detail and discuss its consequences, especially for autoregressive models, as their direct sampling procedure is not compatible with the symmetrization procedure that we found to be optimal.
\end{abstract}

\maketitle  

\section{Introduction}
\label{sec:intro}
Many intriguing phenomena in condensed matter physics emerge in strongly interacting many-body systems, which rarely allow for analytical solutions. Theoretical attempts to understand a problem of interest therefore usually rely on numerical techniques. However, the exponential scaling of required computational resources with the number of system constituents implies that true many-body settings involving hundreds of qubits or more are not amenable to naive classical computational approaches \cite{Feynman1982}.

Therefore, variational approaches have become popular \cite{Schollwoeck2011, Orus2014, Carleo2017} aiming to efficiently parameterize the physically relevant ``corner'' of Hilbert space \cite{Hastings2006, Hastings2007, Eisert2010}. Such approaches reduce the number of required parameters to be sub-exponential, circumventing the curse of dimensionality, at the expense of generality. Various methods have been developed that fall into this category. Tensor network states (TNS) \cite{Schollwoeck2011, Orus2014} and neural quantum states (NQS) \cite{Carleo2017} are particularly popular and important in the case of many-body spin systems, because both are versatile and numerically exact in the sense that the accuracy can be systematically controlled with self-consistent convergence checks. Tensor network approaches comprise many different ansätze \cite{Zwolak2004,Vidal2014,Vidal2007,Shi2006,Cincio2008,Verstraete2004,Klumper1993,Fannes1992}, which are generally composed of a number of tensors with a specific bond dimension, that regulates their expressivity. Matrix product states (MPS) \cite{Schollwoeck2011} form the best studied example of TNS, and allow to encode weakly entangled states in one dimension with great success, especially in conjunction with the celebrated density matrix renormalization group (DMRG) algorithm \cite{White1992,Schollwoeck2005}.
MPS constitute an ideal tool to efficiently represent ground states of one-dimensional gapped Hamiltonians, which feature an area law of entanglement \cite{Hastings2007}. Their limitation is set by the exponential growth of the required bond dimension with entanglement entropy and more involved TNS were devised to mitigate this issue \cite{Orus2019}.
While instances of TNS exist which deal with 2D settings \cite{Verstraete2004,Orus2014,Orus2019}, this regime remains particularly challenging. 

Neural quantum states therefore received a lot of attention upon their introduction as they presented a potential remedy to some of the aforementioned shortcomings of TNS, due to several factors. For one, opposed to the DMRG algorithm, the design of NQS and its associated optimization algorithms do not rely on any spatial structure, rendering them ideal candidates for 2D settings \cite{Carleo2017, Schmitt2020}. Secondly, NQS allow for unprecedented flexibility; in principle any network that maps a spin configuration, i.e. a computational basis state, to an associated complex wave function coefficient presents a valid ansatz \cite{HibatAllah2020, Sharir2020, Valenti2022, Schmitt2020, Pescia2022, Carrasquilla2021a, Carrasquilla2019}. 
Finally, it has been demonstrated that NQS can indeed encode volume-law entangled states without introducing exponential cost \cite{Sharir2021, Deng2017, Gao2017, Passetti2022} in stark contrast to TNS approaches. This observation serves as a strong motivation for the further exploration of NQS
and its yet undetermined limitations.

In this work we aim to shed light on the latter question, focusing mainly on the differences introduced by different network architectures and symmetrization schemes. To this end, we optimize various networks that have been proposed for the use as NQS ansatz functions to represent the ground state of prototypical 2D spin Hamiltonians. In order to connect to the existing literature, we choose the $12\times12$ transverse-field Ising model (TFIM)  \cite{HibatAllah2020, Sharir2020} and $J_1$-$J_2$ model \cite{Nomura2021, Bukov2021, Choo2019} at lattice size $6\times6$ and $10\times10$.

The manuscript is organized as follows: At first we give an introduction to variational Monte Carlo (VMC) and NQS, by introducing the employed optimization strategy (Sec.~\ref{sec:VMC}), the different architectures we benchmarked (\ref{sec:NQS}), as well as the different symmetrization options (Sec.~\ref{sec:symm}). We then present the results on the 2D TFIM as well as the 2D $J_1$-$J_2$ model, where we put an emphasis on the interplay of the symmetrization strategy and the learned sign-structure (Sec.~\ref{sec:results}) and discuss implications for autoregressive networks (Sec.~\ref{sec:conclusion}).

\section{Variational Monte Carlo} 
\label{sec:VMC}

In variational Monte Carlo one paramaterizes a wavefunction as 
\begin{align}
    \ket{\psi_{\vec\theta}}= \sum_{\vec s}\psi_{\vec\theta}(\vec s)\ket{\vec s}\ ,
\end{align}
where $\vec s=(s_1,\ldots,s_N)$ labels the computational basis states of the system composed of $N$ degrees of freedom and ${\vec\theta}$ denotes the vector of variational parameters. Using this ansatz function, one can estimate the expectation values of operators $\hat{O}$ according to
\begin{equation}
    \begin{split}
        \langle \hat{O} \rangle &= \bra{\psi_{\vec\theta}}\hat{O}\ket{\psi_{\vec\theta}} = \sum_{\mathbf{s}\mathbf{s}'} \psi_{\vec\theta}(\mathbf{s})^* O_{\mathbf{s}\mathbf{s}'} \psi_{\vec\theta}(\mathbf{s}')\\
        &= \sum_{\mathbf{s}\mathbf{s}'} |\psi_{\vec\theta}(\mathbf{s})|^2 O_{\mathbf{s}\mathbf{s}'} \frac{\psi_{\vec\theta}(\mathbf{s}')}{\psi_{\vec\theta}(\mathbf{s})}
        = \left\langle \sum_{\mathbf{s'}} O_{\mathbf{s}\mathbf{s}'} \frac{\psi_{\vec\theta}(\mathbf{s}')}{\psi_{\vec\theta}(\mathbf{s})} \right\rangle
    \end{split}    
\end{equation}
where $\langle \cdot \rangle$ denotes the Monte-Carlo (MC) average with respect to the probability distribution  $|\psi_{\vec\theta}(\mathbf{s})|^2$ from here on.
The local estimator $O_{loc}(\mathbf{s})=\sum_{\mathbf{s'}} O_{\mathbf{s}\mathbf{s}'} \psi_{\vec\theta}(\mathbf{s}')/\psi_{\vec\theta}(\mathbf{s})$ can be evaluated efficiently for local operators $\hat{O}$, meaning that the sample size required to reach a certain precision does not scale with the total system size. This renders the estimation of  $\langle \hat{O} \rangle$ efficient \cite{vdNest2011}. For variationally approximating ground states, we can take $\hat{O}$ to be the system's Hamiltonian $\hat{H}$, compute its expectation value and, since we assumed differentiability with respect to ${\vec\theta}$, optimize by gradient descent. A more elaborate approach uses information about the local curvature of the variational manifold, as measured by the Fubini-Study metric, in form of the quantum geometric tensor $S=\llangle \Gamma_k^*(\mathbf{s}) \Gamma_{k'}(\mathbf{s}) \rrangle$, where $\Gamma_k(\mathbf{s})=\partial_{{\vec\theta}_k} \log\psi_{\vec\theta}(\mathbf{s})$ and $\llangle AB\rrangle=\langle AB \rangle - \langle A \rangle\langle B\rangle$. This results in the commonly used stochastic reconfiguration (SR) algorithm \cite{Sorella2007}, in which the parameter update rule is given by
\begin{equation}
\label{eqn:tdvp_update}
    {\vec\theta}_k^{(n+1)} = {\vec\theta}_k^{(n)} - \tau \sum_{k'} \Re(S^{-1})_{kk'}F_{k'} |_{{\vec\theta}={\vec\theta}^{(n)}}.
\end{equation}
Here, $F_{k}=\partial_{{\vec\theta}_{k}}\langle E_{loc}(\mathbf{s})\rangle=2\Re(\llangle \Gamma_{k}^*(\mathbf{s})E_{loc}(\mathbf{s}) \rrangle$) and $\tau$ is the update step size. As $S$ can be rank-deficient, regularization techniques need to be applied upon computing the inverse \cite{Carleo2017, Schmitt2020}. In this work, we achieve regularization by scaling all diagonal entries of $S$ by a factor of $1+\delta_1$ and adding an identity matrix scaled with $\delta_2$ to $S$. During optimization a second order Runge-Kutta integrator with adaptive (imaginary) time step $\tau$ is employed to obtain the integrated evolution of the network parameters. Further details are given in Appendix~\ref{appendix:SR}.

\section{Neural Quantum States}
\label{sec:NQS}
Neural quantum states form a particular class of functions that can be used as an ansatz, $\psi_{\vec\theta}$, in the VMC framework described in Sec.~\ref{sec:VMC}. In this case, the ansatz function is given by an artificial neural network (ANN), which defines a non-linear differentiable map from spin configurations $\mathbf{s}$ to the associated (generally complex) wave function coefficient $\psi_{\vec\theta}(\mathbf{s})$.In fact, it is common practice to have the ANN produce the logarithmic wave function coefficient $\chi_{\vec\theta}(\vec s)$ such that $\psi_{\vec\theta}(s)\equiv\exp(\chi_{\vec\theta}(\vec s))$ can accurately capture coefficients over multiple orders of magnitude. Typically, neural networks are built up in layers which iteratively transform the input to a desired output, constituting a very general class of function approximators that are particularly attractive due to their flexibility in construction and the existence of universal approximation theorems, stating that neural networks can approximate any function given sufficiently many parameters \cite{Cybenko1989,Hornik1991,Kim2003,LeRoux2008}.

Various architectures have been proposed for the use in NQS applications, but a broad comparative benchmark within a unified study is so far missing. We attribute this to both the technical difficulties of implementation and the many details concerning the structure and optimization of the networks, which can have a strong effect on performance. Especially the choice of hyperparameters as well as architecture details, e.g. whether amplitude and phase are treated by separate networks or in a unified scheme, can be challenging. 

We design the networks to encode both phase and amplitude simultaneously, except for the case of the TFIM where no phase is modelled since the ground state is known to be positive \cite{Bravyi2007}. When the network models both phase and amplitude, we design the feed-forward based architectures as holomorphic maps, using complex parameters and outputting a single complex number $\chi_{\vec\theta}(\mathbf{s})$.
The recurrent architectures, in contrast, do not define holomorphic maps and therefore utilize real parameters.
For detailed explanations on the utilized networks see Appendix~\ref{appendix:ANNs}.

Within this work we consider the following network architectures:

\textbf{Restricted Boltzmann Machines} -- The earliest NQS architectures relied on a dense single-layer feed-forward network, usually referred to as restricted Boltzmann machine (RBM) in the NQS context \cite{Carleo2017, Czischek2018, Sun2022, Deng2017, Nomura2021, Viteretti2022, Hofmann2022}. Due to the dense connectivity, the spatial structure of the input is not natively represented by the network architecture. To add a notion of locality to the RBM ansatz one can include the product of physically coupled spins as input features to obtain a correlator RBM (CorrRBM), as proposed in \cite{Valenti2022}.

\textbf{Convolutional Neural Networks} -- A further generalization of RBMs are convolutional neural networks (CNN) \cite{Schmitt2020, Choo2019, Schmale2022, Gutierrez2022}. The layers in deep CNNs are defined by a number of filters of a certain width, thereby allowing great flexibility in the design of the network. By choosing the depth, i.e. the number of layers to one, and the filter size such that it spans the entire system one obtains a translationally invariant RBM, meaning that symmetrized RBMs are a strict subclass of CNNs. By using smaller filters and multiple layers, an intrinsic representation of locality in the network is restored based on a hierarchical representation of features, loosely reminiscent of tree tensor networks (TTN) \cite{Shi2006} or the multiscale entanglement renormalization ansatz (MERA) \cite{Cincio2008}. In analogy to the CorrRBM, adding correlations of the spin configurations to the input extends the CNN to a correlator CNN (`CorrCNN').

\textbf{Recurrent Networks.} A very different form of ansatz is obtained by substituting the feed-forward based network with a recurrent architecture, which assigns the coefficient $\psi_{\vec\theta}(\mathbf{s})$ as a product $\psi_{\vec\theta}(\mathbf{s}) = \psi_{\vec\theta}(s_1) \cdot \psi_{\vec\theta}(s_2|s_1) \cdot ... \cdot \psi_{\vec\theta}(s_N|s_{N-1}...s_1)$ \cite{Sharir2020, HibatAllah2020, Reh2021, Luo2022, Donatella2022, Vicentini2022}. A central motivation for this construction is the fact that such architectures allow for autoregressive sampling. This means that a new independent sample  can be generated with a single network evaluation, such that no Markov chain Monte Carlo (MCMC) with potentially long autocorrelation or thermalization times is required. A recurrent architecture is defined by a cell containing the variational parameters. It is then iteratively scanned over the input, i.e. the spin configuration $\mathbf{s}=(s_1,\dots,s_N)$. Since it is always the same cell that is scanned over the input sequence, the number of variational parameters does not grow with the system size; this can enhance efficiency, but, on the other hand, the inherently sequential design does not allow for parallelization. The architecture of the recurrent cell is highly customizable; we refer to the vanilla recurrent neural-network with a single layer as `RNN', and use the abbreviations `LSTM' and `GRU' for the long short-term memory \cite{Hochreiter1997} and gated recurrent unit \cite{Cho2014}, respectively.

\section{Symmetries}
\label{sec:symm}
When the Hamiltonian describing the spin system exhibits certain symmetries, such as invariance under translations, rotations, reflections or parity symmetry, its ground state exhibits the same symmetries. This a-priori knowledge of properties that the ground state wave-function must fulfill, can be used to restrict the variational optimization to the correct symmetry sector and thereby to enhance the performance of the algorithm. 
Since not all of the considered network architectures can be composed of equivariant layers to incorporate symmetry \cite{Bronstein2021},
we instead follow the common approach to symmetrize the wave-function by averaging procedures of all symmetry equivalent configurations as is commonly done \cite{Nomura2021, Schmitt2020, HibatAllah2020}. The symmetrized coefficient $\psi_{\vec\theta}^S(\textbf{s})$ will then read $\psi_{\vec\theta}^S(\textbf{s})=\mathrm{Avg}\{\psi_{\vec\theta}(\sigma(\mathbf{s}))|\sigma\in\mathcal{S}\}$ with $\mathcal{S}$ the set of all symmetry operations and Avg denoting one of the three symmetrization methods listed below.
In fact, we will show that the details of this symmetrization procedure are crucial for the performance of the algorithm and that not all architectures are amenable to the symmetrization procedure we found to be optimal.
We differentiate between three different procedures for defining symmetrized wave function coefficients $\psi_{\vec\theta}^S(\mathbf{s})$ from the network outputs $\chi_{\vec\theta}(\mathbf{s})$, 

\begin{itemize}
    \item
    \textbf{Bare-Symmetry:}
        \begin{equation}
            \label{eqn:bare_symmetry}
            \psi_{\vec\theta}^S(\mathbf{s}) = \exp\left(\frac{1}{|\mathcal{S}|} \sum_{\mathbf{s'} \in \mathcal{S}(\mathbf{s})} 
            \chi_{\vec\theta}\big(\mathbf{s'}\big)\right)
        \end{equation}
    \item
    \textbf{Exp-Symmetry:}
    \begin{equation}
        \psi_{\vec\theta}^S(\mathbf{s}) = \frac{1}{|\mathcal{S}|} \sum_{\mathbf{s'} \in \mathcal{S}(\mathbf{s})} 
        \exp\Big(\chi_{\vec\theta}\big(\mathbf{s'}\big)\Big)
        \label{eqn:exp_symm}
    \end{equation}
    \item
    \textbf{Sep-Symmetry:}
    \begin{equation}
    \label{eqn:sep_symmetry}
    \begin{split}
        \psi_{\vec\theta}^S(\mathbf{s})=& \sqrt{\frac{1}{|\mathcal{S}|} \sum_{\mathbf{s'} \in \mathcal{S}(\mathbf{s})} 
        \exp\Big(2\text{Re}\Big[\chi_{\vec\theta}\big(\mathbf{s'}\big)\Big]\Big)} \\ 
        &\times
        \exp\left(i\arg\left(\sum_{\mathbf{s'} \in \mathcal{S}(\mathbf{s})} \exp\left(i\ \text{Im}\Big[\chi_{\vec\theta}\big(\mathbf{s'}\big)\Big]\right)\right)\right)
    \end{split}
    \end{equation}
\end{itemize}

Eq.~\eqref{eqn:exp_symm} is a natural choice when the ANN output is considered to be the logarithmic wave function coefficient.
Another option to proceed is to exponentiate the logarithmic coefficients prior to averaging, as done in Eq.~\eqref{eqn:exp_symm}, resulting in a symmetrization procedure with the potential that phases may positively or negatively interfere, as opposed to Eq.~\eqref{eqn:bare_symmetry} and Eq.~\eqref{eqn:sep_symmetry}. Finally, the symmetrization procedure in Eq.~\eqref{eqn:sep_symmetry} is designed to be compatible with an autoregressive property, which is lost when choosing one of the other options. The reason for this is that the relation 
\begin{equation}
\sum_{\mathbf{s'} \in \mathcal{S}(\mathbf{s})}|\psi_{\vec\theta}^S(\mathbf{s'})|^2= \sum_{\mathbf{s'} \in \mathcal{S}(\mathbf{s})} |\psi_{\vec\theta}(\mathbf{s'})|^2.
\label{eqn:autoreg_sym_requirement}
\end{equation}
has to be fulfilled for direct sampling.
The only way to generate a new configuration from the distribution encoded in an autoregressive network is to sequentially sample the local configurations $s_i$ from the conditional probabilities $|\psi_{\vec\theta}(s_i|s_{i-1}...s_1)|^2$. Following this procedure, symmetry-equivalent configurations will in general not be generated with identical probability. However, if Eq.~\eqref{eqn:autoreg_sym_requirement} holds, the frequency of samples from one equivalence class matches its probability given by the symmetrized ansatz $\psi_{\vec\theta}^S(\vec s)$, which is sufficient for a representative set of samples.

All presented symmetrization options have been applied in previous works: The bare option was for example used in \cite{Schmitt2020}, the exponential option in \cite{Nomura2021} and the separate option in \cite{HibatAllah2020}. Our contribution is a direct comparison between these options, applied to the same physical problem with otherwise identical network architectures.

\section{Results}
\label{sec:results}
In the following we will examine the performance of different networks on the TFIM and $J_1$-$J_2$ model on a two-dimensional square lattice.
Since we want to learn about the distinct representational power of the networks, they are designed to have approximately the same number of parameters, see Table \ref{tab:networkSpecifications} in Appendix~\ref{appendix:net_specs}. Note that this can imply different computational cost for the different networks, as for example the evaluation of autoregressive architectures is more demanding than for a feed-forward architecture. We trained the networks until convergence was reached, irrespective of computational cost. Details of the network architectures can be found in Appendix~\ref{appendix:ANNs}.

\subsection{Transverse field Ising model}
The transverse field Ising model
\begin{equation}
\hat{H} = -J\sum_{\langle ij \rangle} \hat{\sigma}^i_z \hat{\sigma}^j_z - h \sum_i \hat{\sigma}^i_x
\end{equation}
on a 2D square lattice features an Ising interaction between nearest neighbors and an external field in $x$-direction, giving rise to a quantum phase transition from a ferromagnetic to a paramagnetic phase at $h=h_c\approx3.044J$ \cite{Bloete2002}.
The stoquastic nature of the Hamiltonian \cite{Bravyi2007} make it an ideal benchmark case for various approximative methods, such as quantum Monte Carlo (QMC) and MPS. We compare to both techniques with data taken from \cite{HibatAllah2020} and \cite{Sharir2020}. 

The stoquasticity implies a positive ground-state wave function, meaning that the NQS networks do not need to encode a complex phase. Here, we limit ourselves to the bare symmetry option. We employ open boundary conditions, to allow for comparisons with \cite{Sharir2020, HibatAllah2020}. 

\begin{figure}[t!]
    \includegraphics[width=\linewidth]{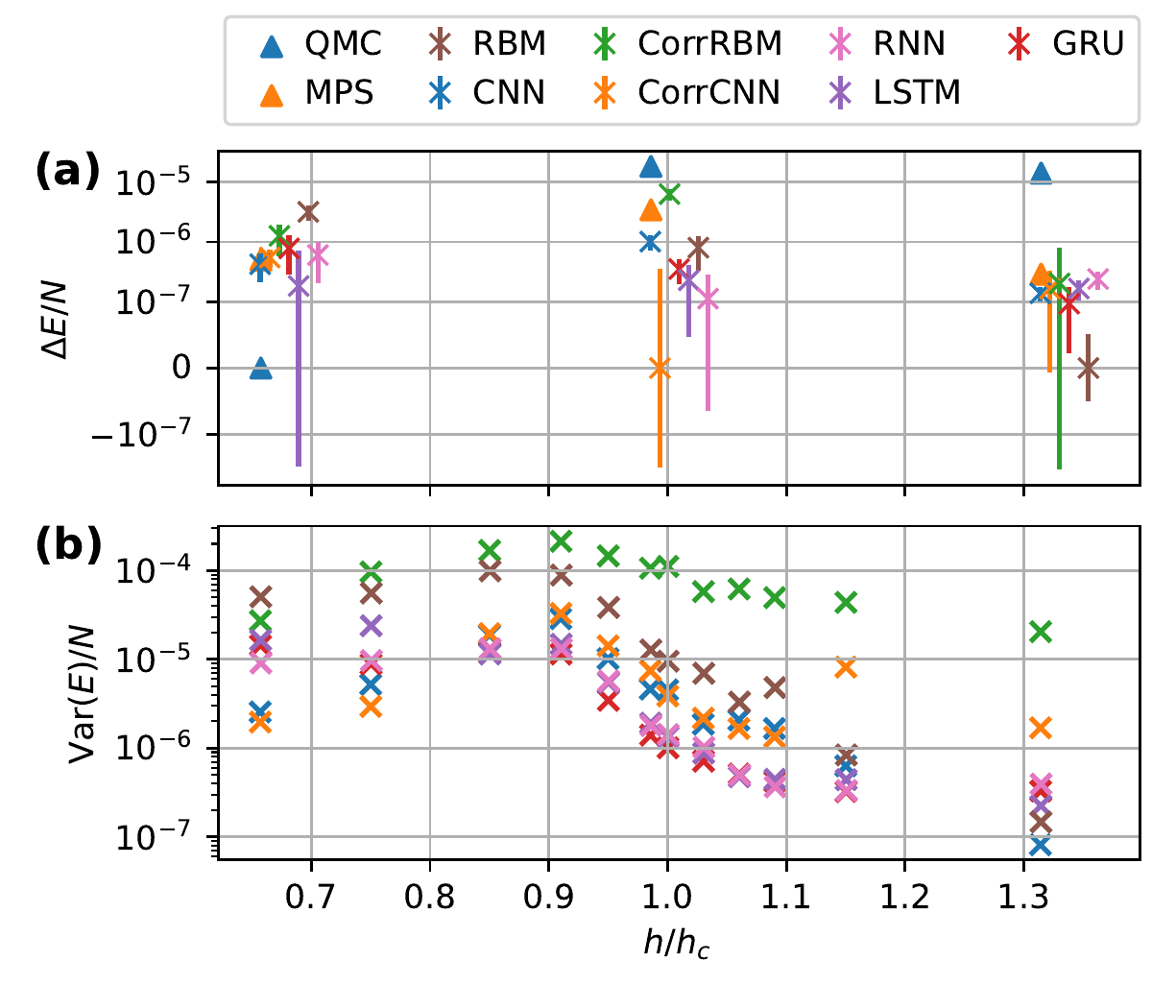}
    \caption{Performance of the tested network architectures for finding the ground state of the 2D $12\times 12$ TFIM at different magnetic field strengths. Here all networks use the bare-symmetrization as no phase is modelled. In (a) the deviation from the lowest observed energy density is shown for the magnetic fields $h\in [2, 3, 4]$. The data points are artificially shifted horizontally such that differences are more easily visible. In (b) the energy variance per spin is plotted, which serves as an additional performance indicator. The QMC and MPS data was taken from \cite{HibatAllah2020} and \cite{Sharir2020}.}
    \label{fig:TFIM_QPT}
\end{figure}

We present our results in Fig.~\ref{fig:TFIM_QPT}. Panel (a) shows the deviation of the predicted ground state energy density from the lowest value attained by the various networks and QMC and MPS at each field strength $h\in [2, 3, 4]$. In order to ease the legibility we shifted the field-strenghts slightly. In addition, since no exact numerical benchmark value for the energy itself can be given, we  consider the variance of the energy estimate as a performance indicator, that vanishes when the network represents an eigenstate of the system, such as the ground-state. Figure~\ref{fig:TFIM_QPT}(b) shows the energy variance as a function of field strength for the various network architectures. The first observation is that for all architectures, performance depends significantly on the strength of the magnetic field. In the region of the critical value performance diminishes as one might expect, as correlations grow longer ranged. The worst performance is however found at $h=0.91 h_c$ slightly away from the critical point, which might be an indicator of finite-size effects being present. For the differences between the networks it is more difficult to make out clear trends. While the RBM and correlator RBM architecture seem to give the worst performance, CNN and correlator CNN perform better, which indicates that increasing the depth of the network at the expense of smaller filter sizes is beneficial, especially in the ferromagnetic phase. Autoregressive nets perform reasonably well for all values of $h$, especially in the paramagnetic regime where they outperform the other architectures. Interestingly, the details of the recurrent cell do not seem to have an influence in this scenario as RNN, LSTM and GRU show almost identical performance.

We conclude that the ground state of the stoquastic TFIM can be approximated well, without major differences among network architectures.
In the following, we therefore shift our attention to the 2D $J_1$-$J_2$ model, i.e. a non-stoquastic Hamiltonian with a non-trivial sign structure. At the point $J_2=0.5J_1$ the ground state is maximally frustrated, giving rise to exotic phases such as a spin-liquid and valence bond solid state \cite{Nomura2021}, which are characterized by high entanglement and the absence of an energy gap.

\subsection{\texorpdfstring{$\boldsymbol{J_1-J_2}$ model}{$J_1$-$J_2$ model}}

\begin{figure*}[t]
    \centering
    \includegraphics[width=\textwidth]{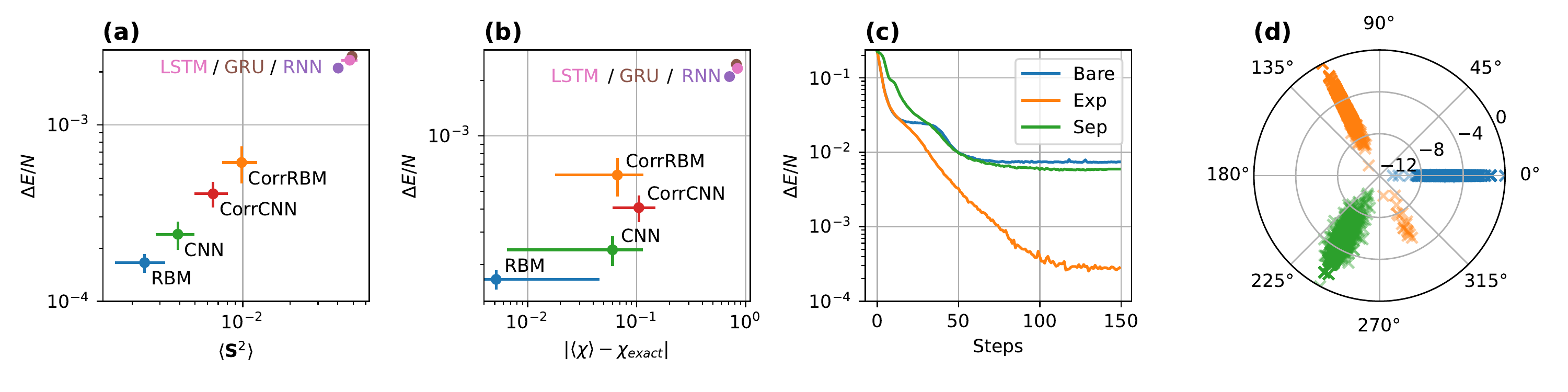}
    \caption{Results on the 2D $6\times 6$ $J_1$-$J_2$ model. In (a) and (b) the performance of the different architectures is depicted as a function of observables of interest. The errorbars which are smaller than the filled in circles are omitted. In (c), the effect of the different symmetrization procedures is shown for the case of the RBM architecture, leaving all other specifications of the network unchanged. In (d), sampled wave function coefficients of the network are depicted in the complex plane for the symmetrizations shown in (c), with a logarithmic scale for the absolute value. Only in the exponential case were we able to observe a non-trivial sign structure with deviations from the Marshall sign rule, in which there were samples on the opposite site of the bulk.}
    \label{fig:J1J2}
\end{figure*}

The $J_1$-$J_2$ model features competing antiferromagnetic Heisenberg couplings between nearest- and next-nearest neighbours, with respective strengths $J_1$ and $J_2$, such that 
\begin{equation}
\hat{H} = J_1 \sum_{\langle ij \rangle} \hat{\boldsymbol{\sigma}}_i \cdot \hat{\boldsymbol{\sigma}}_j + J_2 \sum_{\llangle ij \rrangle} \hat{\boldsymbol{\sigma}}_i \cdot \hat{\boldsymbol{\sigma}}_j\ .
\end{equation}
It has been the focus of numerous works, especially in the context of neural quantum states \cite{Nomura2021, Nomura2021b, Choo2019, Liang2018, Ferrari2019} due to its challenging nature that is characteristic for 2D frustrated ground states.
Depending on the relative strength of interactions, one finds the system in the Néel ($J_2 \lesssim 0.49 J_1$) or striped ($J_2 \gtrsim 0.61 J_1$) phase, while a spin-liquid and valence bond solid phase exists between the two \cite{Nomura2021b}. All of our experiments are carried out at the point $J_2/J_1=0.5$, where frustration is maximal and we employ periodic boundaries. The abundance of symmetries reduces the size of the relevant Hilbert space sector, allowing to exactly compute the ground-state energy for a lattice of size $6\times6$, which serves as a reference value. 

Due to the non-stoquastic nature of the Hamiltonian, the ground state is not positive and therefore a major challenge consists in finding the ground state sign structure \cite{Westerhout2020,Szabo2020,Bukov2021}. At $J_1=0$ or $J_2=0$, the phase follows the Marshall-Peierls sign rule, which states that the sign of each coefficient $\psi(\mathbf{s})$ is $(-1)^{N_{\uparrow\in A}(\mathbf{s})}$, where $A$ denotes one of the two sublattices and $N_{\uparrow\in A}(\mathbf{s})$ is the number of up spins in configuration $\mathbf{s}$ on sublattice $A$. When going to finite ratios $J_2/J_1$, the Marshall-sign rule does no longer hold exactly, but presents a good approximation. Hence, we choose to hard-code the sign rule in form of a transformed Hamiltonian, which is also explained in Appendix~\ref{appendix:signrule}, and let the network learn deviations from it. We therefore change the design of the feed-forward based architectures to holomorphic maps with complex network parameters, while the autoregressive architectures stay non-holomorphic maps using real parameters but are allowed to model a phase. 

Additionally, the ground state is known to lie in the zero magnetization sector, implying that the wave function coefficients are only non-zero for configurations $\mathbf{s}$ with the same number of up and down spins. We exploit this by only constructing sample configurations that fulfill this condition.

As was demonstrated in \cite{Nomura2021}, supplying the network with physical information in form of symmetries improves performance dramatically. We therefore symmetrize our ansatz function with all four present symmetries, i.e. translations, the $C_4$ point group and the $\mathbb Z_2$ spin-flip symmetry. As described previously, different options regarding the details of symmetrization exist, and the data presented below reveals that these details have significant effect on the performance of the algorithm.

Let us, however, first examine the performance differences between the different network architectures, shown in Fig.~\ref{fig:J1J2}(a) and (b).
We show the deviation of the energy density from the exact ground state value as a function of the total magnetization
\begin{equation}
    \hat{\mathbf{S}}^2 = \sum_{ij} \hat{\boldsymbol{\sigma}}_i \cdot \hat{\boldsymbol{\sigma}}_j
\end{equation}
and as a function of the structure factor
\begin{equation}
    \hat{\chi} = \sum_{ij} \hat{\boldsymbol{\sigma}}_i \cdot \hat{\boldsymbol{\sigma}}_j e^{i\mathbf{q}\cdot(\mathbf{r}_i-\mathbf{r}_j)}
\end{equation}
for $\mathbf{q} = (\pi, \pi)$, where the sum runs over all pairs of lattice sites. The choice of these observables is physically motivated: The Hamiltonian's $SU(2)$ symmetry implies that the ground-state fulfills $\langle \hat{\mathbf{S}}^2\rangle = 0$, and thus the deviation from zero serves as another figure of merit for the approximation quality. The structure factor is a crucial obvservable providing insight into the magnetization structure, i.e. whether the striped or Néel phase is observed. The observables are estimated using $10^4$ samples. The error bars shown in the figure are obtained by estimating the observables based on $10$ independently drawn sample sets and determining the fluctuations between them. The feed forward architectures, employing the exponential symmetrization strategy, perform significantly better compared to the autoregressive networks using the separate (sep-)symmetrization strategy. In Fig.~\ref{fig:J1J2}(c) we compare the effect of the three different strategies for the RBM. The exponential option, utilized in \cite{Nomura2021}, performs best by far. Other strategies, that were for example used in \cite{Schmitt2020} or \cite{HibatAllah2020} perform considerably worse for the given task. 

In a next step, we provide an analysis revealing the origin of the observed performance differences. 
Concretely, we examine the capability of the different symmetrization methods to capture deviations from the Marshall sign rule. In Fig.~\ref{fig:J1J2}(d) we show argument and amplitude of the complex wave function coefficients that are obtained by sampling the NQS; notice that all coefficients from one wave function can be rotated by an arbitrary angle in the polar plot, corresponding to an irrelevant global phase. Importantly, deviations from the Marshall sign rule are only observed in the case of exp-symmetrization, while the other methods only optimize the wave function amplitudes, thus not being able to reveal the non-trivial physics in this problem. This explains the inferior performance of the alternative symmetrization schemes observed in Fig.~\ref{fig:J1J2}(c). This, of course, does not mean that we cannot model such behaviour with the other strategies -- it can, however, hint at the fact, that finding such sets of variational parameters using the stochastic reconfiguration algorithm is challenging. 

\begin{figure}[t]
    \includegraphics[width=.9\linewidth]{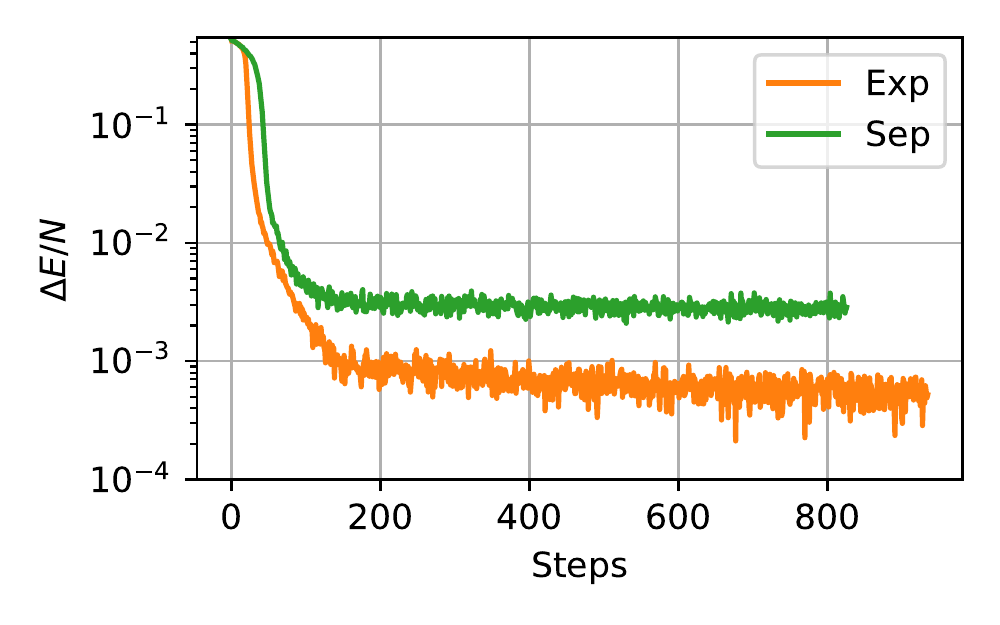}
    \caption{Learning curve of an RNN with the sep-symmetry and exact samples (green) compared to the same network using the exponential symmetrization strategy with MCMC samples (orange). As is clearly visible, the network profits greatly from the exponential symmetrization strategy which is however incompatible with exact sampling, thereby taking away the major benefit of autoregressive architectures.}
    \label{fig:J1J2_autoreg}
\end{figure}

We now discuss the implications of this finding. In autoregressive architectures, samples are drawn from the unsymmetrized distribution, since we cannot average over all symmetry invariant configurations while sampling. One therefore needs to ensure that the symmetrized probability measure of a configuration is equal to the non-symmetrized one, ruling out the exponential symmetrization strategy in conjunction with autoregressive samples, since it violates Eq.~\eqref{eqn:autoreg_sym_requirement} and similarly for the bare symmetrization. To verify the influence of the chosen symmetrization on the performance also for recurrent architectures, we compare the performance of a sep-symmetrized RNN optimized using autoregressive sampling to its exp-symmetrized counterpart with MCMC sampling in Fig.~\ref{fig:J1J2_autoreg}. The performance difference again reveals that the exponential symmetrization procedure significantly improves the quality of the results, although it prohibits autoregressive samples. 
Therefore, the limited flexibility in choosing the symmetrization scheme constitutes a significant drawback for autoregressive architectures when trying to learn the non-trivial sign-structures of the frustrated ground state.

Having found clear indication that the details of symmetrization influence the results, we now give an intuitive understanding as to why this may be the case. One immediate observation is the fact that only in the exponential case the phase can influence the final amplitude, since the coefficients may positively or negatively interfere, yielding a potential gain in expressivity. Similarly, if the prediction of the phase were random (or noisy) for a given spin configuration, the sum of these noisy coefficients would be small in absolute value. Thereby, it seems plausible that in the case of bad generalization of the ANN, the resulting coefficients are automatically suppressed.
The interplay between complex phases and amplitudes can also affect the optimization: The logarithmic derivatives $\Gamma_k(\mathbf{s})=\partial_{\theta_k}\log|\psi_{\vec\theta}^S|+\partial_{\theta_k}\arg[\psi_{\vec\theta}^S]$ of the symmetrized coefficients consist of the sum of the gradients of the (log-)amplitude and the complex phase. If only the amplitudes of the non-symmetrized coefficients $\psi_{\vec\theta}(\vec s)$ contribute to the symmetrized amplitude, as is the case for the bare- and sep-symmetrization, its gradients with respect to the complex phases of $\psi_{\vec\theta}(\vec s)$ vanish. This is in contrast to the exp-symmetry case, where the gradients of the symmetric amplitude with respect to the non-symmetric phases generally take finite values, since they influence the symmetrized amplitude.

\subsection{Large system sizes}
Having gained insight into network specific advantages and disadvantages, we finally want to answer the question which network architectures can be deemed suitable to scale up to larger system sizes. To consider an architecture scalable we demand that both the number of its parameters and the network evaluation cost should at most grow mildly with system size $N$.
The RBMs and its variants violate the first requirement as the number of parameters grows quadratically given a constant ratio $\alpha$ between the number of hidden and visible neurons. For recurrent networks, one finds a quadratic scaling of the evaluation cost, as the cost of evaluating a single (unsymmetrized) configuration grows linear with system size, since the cell needs to be scanned over longer inputs, while symmetrization gives another factor $N$. 

This leaves CNNs as the only viable option, as they allow for sub-quadratic growth of parameters using small filter sizes in deep architectures, still allowing for system wide correlations, and evaluations that approximately scale linear in system size since the filters need to be scanned over larger inputs. Since we saw in Fig.~\ref{fig:J1J2} that adding correlation input features does not add to performance, we test the CNN on a $10 \times 10$ lattice and compare to previous works in Table~\ref{tab:J1J210x10}. 
All listed energies stem from NQS approaches, where those that used physically informed input in addition to the NQS are marked with a `+'. The additional input is using variational parameters that define the weight of the projection of the basis state onto a physically motivated state, such as a pair state in the pair-projection (PP) algorithm or a Gutzwiller-projected (GP) state. The performance of the CNN presented here is comparable to \cite{Ferrari2019}, without using any physically informed input.

\begin{table}[]
    \centering
    \begin{tabular}{c|c|c}
         Architecture & Energy & Reference  \\
         \hline
         CNN & -0.473591 & \cite{Liang2018} \\
         CNN & -0.49516(1) & \cite{Choo2019} \\
         RBM + GP & -0.49575(2) & \cite{Ferrari2019} \\
         \textbf{CNN} & \textbf{-0.49586(4)} & \textbf{This Work} \\
         RBM + PP & -0.497629(1) & \cite{Nomura2021b} \\
    \end{tabular}
    \caption{Ground state energies achieved with neural network quantum states on the $10 \times 10$ $J_1$-$J_2$ model with periodic boundary conditions at the point $J_2/J_1 = 0.5$. References \cite{Ferrari2019} and \cite{Nomura2021b} incorporate projection procedures, thereby providing physical information beyond symmetries which are not present in the other approaches.}
    \label{tab:J1J210x10}
\end{table}

\section{Conclusion and Outlook}
\label{sec:conclusion}
In this work, we systematically compared the performance of different network architectures and symmetrization strategies when trying to find ground states of many-body spin Hamiltonians. While there are no clear trends visible in the case of the 2D TFIM model, the disparities in the $J_1$-$J_2$ model are pronounced, where the feed-forward architectures clearly outperform the autoregressive architectures. We identified the details of the symmetrization strategy as a key to higher performance, which hinders learning the correct sign structure if not done correctly. Specifically, it is imperative to add the wave function coefficients of all symmetry-invariant configurations such that they may interfere positively or negatively as layed out in Eq.~\eqref{eqn:exp_symm}. Surprisingly, it is therefore insufficient to `only' search in the correct symmetry sector.

This finding has direct implications for autoregressive architectures, since their sampling strategy is not amenable to the exponential symmetrization strategy described in Eq.~\eqref{eqn:exp_symm}, as it violates Eq.~\eqref{eqn:autoreg_sym_requirement}. We make this explicit in Fig.~\ref{fig:J1J2_autoreg}, by showing that an autoregressive model using MCMC samples outperforms the same model with exact samples if the symmetrization strategy is such that interference among the coefficients is possible. Therefore, learning quantum states which feature a non-trivial sign structure appears at least challenging with autoregressive models, if symmetrization is required at the same time. 

We conjecture, that further progress will rely on symmetrical wavefunction ansätze that support good scaling characteristics to larger system sizes. The symmetry invariant map should add up the generated coefficients using the exp-symmetry operation allowing for readily learning a non-trivial sign structure as opposed to the other cases. We have found the CNN-architecture to be suitable for this purpose and demonstrated comparable performance to previous works, which amended NQS wave functions with additional physical bias.

\textit{Note added:} During the preparation of this manuscript Ref.~\cite{Roth2022} appeared, which also studies the $J_1$-$J_2$ model on the square lattice using group convolutional neural networks (GCNNs). The findings of that work support our claim that the exponential operation prior to summation is crucial for optimizing performance.

\textit{Code \& Data availability.}
The code developed for this project relies on the jVMC library \cite{jvmc}, that can be found on \href{https://github.com/markusschmitt/vmc_jax}{GitHub: markusschmitt/vmc\_jax}. Data is available upon request.

\acknowledgments
The authors would like to thank Or Sharir, Mohamed Hibat-Allah and Francesco Ferrari for providing data from their respective works. 

This work is supported by the Deutsche Forschungsgemeinschaft (DFG, German Research Foundation) under Germany’s Excellence Strategy EXC2181/1-390900948 (the Heidelberg STRUCTURES Excellence Cluster) and within the Collaborative Research Center SFB1225 (ISOQUANT). This work was partially financed by the Baden-Württemberg Stiftung gGmbH. The authors acknowledge support by the state of Baden-Württemberg through bwHPC
and the German Research Foundation (DFG) through Grant No INST 40/575-1 FUGG (JUSTUS 2 cluster). The authors gratefully acknowledge the Gauss Centre for Supercomputing e.V. (www.gauss-centre.eu) for funding this project by providing computing time through the John von Neumann Institute for Computing (NIC) on the GCS Supercomputer JUWELS \cite{JUWELS} at Jülich Supercomputing Centre (JSC).

\iftrue
\appendix

\section{Regularization, Inversion and Imaginary Time Evolution}
\label{appendix:SR}
As laid out in the main text, the stochastic reconfiguration algorithm culminates in the parameter evolution equation
\begin{equation}
    \sum_{k'} \Re(S^{-1})_{kk'}\dot{{\vec\theta}}_{k'} = F_{k}.
\end{equation}
Since $S$ can be singular, rendering its inverse ill defined, one needs to regularize $S$ in order to solve for $\dot{{\vec\theta}}$. We achieve this in two steps, first multiplying all diagonal elements with a factor $1+\delta_1$ and then adding $\delta_2$ to all diagonal entries. We choose $\delta_1$ on the order of 10 and let it decay exponentially during optimization, while $\delta_2$ is fixed around $10^{-4}$.
We then find $\dot{{\vec\theta}}$ using a linear solver (we used the \texttt{jax} implementation of the \texttt{scipy} linear solver \texttt{jax.scipy.linalg.solve}) before integrating ${\vec\theta}$ using a second-order Runge-Kutta scheme with adaptive step-size \cite{Schmitt2020}.

\section{Neural Network Architectures}
\label{appendix:ANNs}
\textit{Restricted Boltzmann Machines (RBMs).} A dense, single-layer feedforward network with $N$ input and $M$ hidden neurons is, somewhat imprecisely \cite{Hinton2012}, referred to as RBM in the NQS literature. Its output is the sum of the hidden layer with an activation function applied. Usually, this function is chosen to be the logarithm of the hyperbolic cosine. In the complex plane, this function has poles at $z=i\pi(1/2 + n)$ for integer $n$, which is why we choose to approximate the activation function with its first two non-vanishing Taylor series terms, mitigating this problem. The number of parameters for a network with weights and biases are given by 
\begin{equation}
    n_P=2NM + 2M,
\end{equation} where the factor of 2 arises since all parameters are complex. In varying the number of hidden neurons $M$, one can control the complexity and expressivity of the network. The dense layer bears no physical motivation, as it is irrespective of locality, begging the question for more physically motivated architectures. 

\textit{Convolutional Neural Networks (CNNs).} In contrast to RBMs, CNNs allow to keep a notion of locality, by using filter sizes that are smaller than the system size. These are scanned over the input, automatically respecting translational equivariance. CNNs allow to be made more expressive by tuning the depth ($D$), filter size ($S_F$) and the number of filters/channels ($N_F$). Without biases, the resulting number of parameters is
\begin{equation}
    n_P = 2\sum_{d=1}^D S_F^d N_F^i N_F^{i-1},
\end{equation}
with $N_F^0=1$ for the input spin configuration.
In the end, the output is summed over the channel dimension before exponentiating and summation over the remaining dimensions, in order to implement the exp-symmetrization option.

\textit{Correlation Networks.} In order to guide the network towards the important features of the input configuration, \cite{Valenti2022} proposes to add correlations of coupled spins explicitly to the input of the network. In order to test the proposal, we add a new input for every link between interacting spins and define its value to be the product of the interacting spins in the $\{+1, -1\}$ representation, such that features from aligned (anti-aligned) spins enter as $+1$ ($-1$). As this procedure simply alters the input configuration, it is in principle applicable to any (non-autoregressive) architecture. This is because the autoregressive sampling procedure cannot use correlations between spins that were not sampled yet. The number of parameters the correlation networks use is given by the same formulas as previously, but with $N$ substituted by $N$ + $N_c$, where $N_c$ are the number of couplings between spins in the Hamiltonian.

\textit{Recurrent Neural Networks (RNNs).} Recurrent networks scan a cell $f$ over the input configuration $\mathbf{s}$, while maintaining a memory in a hidden state vector $h$ with length $n_h$. This allows to compute and sample the wavefunction based on conditionals, $\psi_{\vec\theta}(\mathbf{s}) = \psi_{\vec\theta}(s_1) \cdot \psi_{\vec\theta}(s_2|s_1) \cdot ... \cdot \psi_{\vec\theta}(s_N|s_{N-1}...s_1)$ \cite{Sharir2020, HibatAllah2020, Reh2021, Luo2022, Donatella2022, Vicentini2022}. The conditionals are obtained using $h_{t+1} = f(h_t, s_t)$, before computing $\psi_{\vec\theta}(s_{t+1}|s_t...s_1) = g(h_{t+1})$, where $g$ is a dense layer with two outputs that we interpret as $\log(\abs{\psi})$ and $\arg(\psi)$ and from which we compute both $\psi_{\vec\theta}(\uparrow|s_t...s_1)$ and $\psi_{\vec\theta}(\downarrow|s_t...s_1)$. Since the conditional wave functions are normalized to allow for autoregressive sampling, they do not correspond to holomorphic functions, which is why we employ real parameters for all recurrent models to predict amplitude and phase as two distinct outputs of the network. 
In two dimensions, the hidden states $h_t$ and $s_t$ are formed by concatenating the hidden states and spin configurations of the top and right (left) neighbour of the currently considered site, depending on whether the site is located on an even (odd) row \cite{HibatAllah2020}. 

Different architectures have been proposed for the cell $f$. In the easiest case, a 'vanilla' RNN-cell is used (referred to as RNN in the main text), which is defined as a single affine map between inputs and the output with an additional activation function. More complicated cells, such as the LSTM (long short-term memory, \cite{Hochreiter1997}) or GRU (gated recurrent unit, \cite{Cho2014}) cell use additional `gates' that allow to keep or discard information from previous sites. As the number of parameters in these architectures is not easily given, we omit a formula for its calculation at this point and refer the reader to \cite{Hochreiter1997, Cho2014} for further details.

\section{Network specifications}
\label{appendix:net_specs}
We summarize the details of the networks, in particular network sizes and symmetries, for all networks that were used in the paper in Table~\ref{tab:networkSpecifications}.

\section{Marshall sign-rule}
\label{appendix:signrule}
The Marshall sign-rule gives the exact sign for each coefficient $\psi(\mathbf{s})$ in the $J_1$-$J_2$ model for $J_1=0$ or $J_2=0$. As defined in the main text, it assigns the sign $(-1)^{N_{\uparrow\in A}(\mathbf{s})}$, where $A$ denotes one of the two sublattices and $N_{\uparrow\in A}(\mathbf{s})$ is the number of up spins in configuration $\mathbf{s}$ on sublattice $A$. It can equivalently be understood as a gauge-transformation of the Hamiltonian, which then reads
\begin{equation}
    H' = J_1 \sum_{\langle ij \rangle} (-\hat{\sigma}_x^i\hat{\sigma}_x^j -\hat{\sigma}_y^i\hat{\sigma}_y^j + \hat{\sigma}_z^i\hat{\sigma}_z^j) + J_2 \sum_{\llangle ij \rrangle} \hat{\boldsymbol{\sigma}}_i \cdot \hat{\boldsymbol{\sigma}}_j.
\end{equation}
Away from the points $J_1=0$ and $J_2=0$ the sign-rule does no longer hold exactly but presents a good approximation, which we use as a starting point from which we aim to learn the correct deviations. Recovering the physics of the original Hamiltonian from $H'$ is achieved by flipping all $x$- and $y$-correlators with support on both sublattices.

\onecolumngrid
\begin{table*}
\begin{tabular}{|c|c|c|c|c|}
\hline
 Figure / Table & Name & No. of Parameters & Symmetry & Specifications\\ \hline
 \ref{fig:TFIM_QPT} & RBM & 2592 & Bare & $M=18$\\
 \ref{fig:TFIM_QPT} & CNN & 2594 & Bare & $D=2$, $N_F=[24, 2]$, $S_F=[6\times 6$, $6\times 6$]\\
 \ref{fig:TFIM_QPT} & CorrRBM & 2645 & Bare & $M=5$ \\
 \ref{fig:TFIM_QPT} & CorrCNN & 2483 & Bare & $D=2$, $N_F=[20,3]$, $S_F=[7\times 7, 5\times 5]$\\
 \ref{fig:TFIM_QPT} & RNN & 2411 & Sep & $n_h=33$\\
 \ref{fig:TFIM_QPT} & LSTM & 2463 & Sep & $n_h=19$\\
 \ref{fig:TFIM_QPT} & GRU & 2490 & Sep & $n_h=21$\\ \hline
 \ref{fig:J1J2} (a) + (b) & RBM & 3996 & Exp & $M=54$\\
 \ref{fig:J1J2} (a) + (b) & CNN & 3962 & Exp & $D=2$, $N_F=[60, 1]$, $S_F=[4\times4, 4\times4]$\\
 \ref{fig:J1J2} (a) + (b) & CorrRBM & 4060 & Exp & $M=14$\\
 \ref{fig:J1J2} (a) + (b) & CorrCNN & 3872 & Exp & $D=2$, $N_F=[15, 1]$, $S_F=[8\times 8, 8\times 8]$\\
 \ref{fig:J1J2} (a) + (b) & RNN & 4138 & Sep & $n_h=44$\\
 \ref{fig:J1J2} (a) + (b) & LSTM & 4010 & Sep & $n_h=24$\\
 \ref{fig:J1J2} (a) + (b) & GRU & 4160 & Sep & $n_h=27$\\ \hline
 \ref{fig:J1J2} (c) + (d) & RBM & 3996 & Bare & $M=54$\\
 \ref{fig:J1J2} (c) + (d) & RBM & 3996 & Exp & $M=54$\\
 \ref{fig:J1J2} (c) + (d) & RBM & 3996 & Sep & $M=54$\\ \hline
 \ref{fig:J1J2_autoreg} & RNN & 3442 & Exp & $n_h=40$\\
 \ref{fig:J1J2_autoreg} & RNN & 3442 & Sep & $n_h=40$\\
 \hline
 \ref{tab:J1J210x10} & CNN & 10952 & Exp & $D=2$, $N_F=[75, 1]$, $S_F=[6\times 6, 6\times 6]$\\
 \hline
\end{tabular}
\caption{Characteristics of the networks that were utilized throughout the paper.}
\label{tab:networkSpecifications}
\end{table*}
\twocolumngrid
\fi

\bibliography{refs}

\end{document}